\documentclass[prl,superscriptaddress,amsfonts,twocolumn,floatfix]{revtex4}

\usepackage{graphicx,color}
\usepackage{amsmath}
\usepackage{amssymb}
\usepackage{bm}
\usepackage{ulem}

\usepackage{txfonts}

\def\Vec#1{\bm{#1}}
\def\Hc2{H_\mathrm{c2}}
\def\214{\mathrm{Sr_2RuO_4}}
\def\327{\mathrm{Sr_3Ru_2O_7}}
\def\Tc{T_\mathrm{c}}
\def\T0{T_\mathrm{c0}}
\def\muup{\mu}

\begin{document}

\title{Orientation of point nodes and nonunitary triplet pairing tuned by the easy-axis magnetization\\ in UTe$_2$}

\author{Shunichiro Kittaka}
\affiliation{Institute for Solid State Physics, University of Tokyo, Kashiwa, Chiba 277-8581, Japan}
\affiliation{Department of Physics, Chuo University, Kasuga, Bunkyo-ku, Tokyo 112-8551, Japan}
\author{Yusei Shimizu}
\affiliation{Institute for Materials Research, Tohoku University, Oarai, Ibaraki 311-1313, Japan}
\author{Toshiro Sakakibara}
\affiliation{Institute for Solid State Physics, University of Tokyo, Kashiwa, Chiba 277-8581, Japan}
\author{Ai Nakamura}
\affiliation{Institute for Materials Research, Tohoku University, Oarai, Ibaraki 311-1313, Japan}
\author{Dexin Li}
\affiliation{Institute for Materials Research, Tohoku University, Oarai, Ibaraki 311-1313, Japan}
\author{Yoshiya Homma}
\affiliation{Institute for Materials Research, Tohoku University, Oarai, Ibaraki 311-1313, Japan}
\author{Fuminori Honda}
\affiliation{Institute for Materials Research, Tohoku University, Oarai, Ibaraki 311-1313, Japan}
\author{Dai Aoki}
\affiliation{Institute for Materials Research, Tohoku University, Oarai, Ibaraki 311-1313, Japan}
\author{Kazushige Machida}
\affiliation{Department of Physics, Ritsumeikan University, Kusatsu, Shiga 525-8577, Japan}

\date{\today}

\begin{abstract}
The gap structure of a novel uranium-based superconductor UTe$_2$, situated in the vicinity of ferromagnetic quantum criticality,
has been investigated via specific-heat $C(T,H,\Omega)$ measurements in various field orientations.
Its angular $\Omega(\phi,\theta)$ variation shows 
a characteristic shoulder anomaly with a local minimum in $H \parallel a$ at moderate fields 
rotated within the $ab$ and $ac$ planes.
Based on the theoretical calculations,
these features can be attributed to the presence of point nodes in the superconducting gap along the $a$ direction.
Under the field orientation along the easy-magnetization $a$ axis, 
an unusual temperature dependence of the upper critical field at low fields together with a convex downward curvature in $C(H)$ were observed.
These anomalous behaviors can be explained on the basis of a nonunitary triplet state model with equal-spin pairing whose $\Tc$ is tuned by the magnetization along the $a$ axis.
From these results, the gap symmetry of UTe$_2$ is most likely described by a vector order parameter of $\Vec{d}(\Vec{k})=(\Vec{b} + i\Vec{c})(k_b + ik_c)$.
\end{abstract}

\maketitle

Exotic superconductivity arising near ferromagnetic instability has been intensively studied for uranium-based superconductors~\cite{Aoki2019review}, 
such as UGe$_2$~\cite{Saxena2000Nature}, URhGe~\cite{Aoki2001Nature}, and UCoGe~\cite{Huy2007PRL}.
These materials are itinerant ferromagnets but become superconducting even in the ferromagnetic phase.
A remarkable feature is the upper critical field $\Hc2$ exceeding the Pauli-limiting field.
Furthermore, field reentrant (reinforced) superconductivity occurs under high magnetic fields along the hard-magnetization axis in URhGe and UCoGe \cite{Levy2005Science,Aoki2019review,Aoki2009JPSJ}, 
in which spins of Cooper pairs would be polarized along the field orientation or the hard-magnetization axes.
These facts demonstrate that these uranium-based superconductors are promising candidates of spin-triplet superconductors.
The results of NMR measurements suggest that ferromagnetic spin fluctuations play a key role in mediating superconductivity~\cite{Hattori2012PRL,Tokunaga2015PRL}.

Recently, a novel uranium-based superconductor UTe$_2$ has been discovered~\cite{Ran2019Science} and becomes a hot topic in the research field of superconductivity.
Notably, it becomes superconducting at a relatively high $\Tc$ of 1.6~K without showing a clear ferromagnetic transition. 
A first-order metamagnetic transition occurs under a magnetic field $\mu_0H$ at 35~T in $H \parallel b$ with a critical end point at roughly 7 -- 11 K~\cite{Miyake2019JPSJ,Knafo2019JPSJ}.
NMR measurements revealed a moderate Ising anisotropy and suggest the presence of longitudinal magnetic fluctuations along the easy-magnetization $a$ axis above 20~K~\cite{Tokunaga2019JPSJ}.
These facts imply that UTe$_2$ is close to ferromagnetic quantum criticality.
Similar to the other three uranium-based ferromagnets, the formation of spin-triplet Cooper pairing has been indicated 
by small decrease in the NMR Knight shift \cite{Nakamine2019JPSJ} and a large $\Hc2$ exceeding the Pauli-limiting field \cite{Knebel2019JPSJ,Ran2019NatPhys,Aoki2019JPSJ}.
Indeed, superconductivity survives up to an extremely high field of 35~T for $H\parallel b$,
which is destroyed abruptly by the occurrence of a metamagnetic transition~\cite{Knebel2019JPSJ,Ran2019NatPhys}.
Furthermore, reentrant superconductivity arises under $\mu_0H$ beyond 40 T tilted away from the $b$ axis toward the $c$ axis by roughly $30^\circ$~\cite{Ran2019NatPhys}.
These facts demonstrate that parallel spin pairing can be formed in UTe$_2$.
In other words, the vector order parameter is favorably aligned to the plane perpendicular to the $a$ axis (i.e., $\Vec{d} \perp \Vec{a}$) at low fields.

One of the remaining questions for UTe$_2$ is a large residual value of the Sommerfeld coefficient in the superconducting state $\gamma_0$,
which is roughly half of the normal state value $\gamma_{\rm n}$ at $\Tc$.
Whereas a nonunitary spin-triplet state was suggested to explain this feature early on~\cite{Ran2019Science}, 
a magnetic contribution was recently proposed as a possible origin because the entropy balance between superconducting and normal states is not satisfied~\cite{Metz2019PRB}. 
Moreover, a primary question is the gap symmetry which is closely related to exotic pairing mechanisms. 
The presence of linear point nodes in the superconducting gap has been suggested from 
specific heat \cite{Ran2019Science}, nuclear relaxation rate $1/T_1$ \cite{Nakamine2019JPSJ}, penetration depth \cite{Metz2019PRB}, and thermal conductivity~\cite{Metz2019PRB} measurements.
Although the results of recent scanning tunneling microscopy (STM) experiments suggest a chiral order parameter~\cite{Jiao2019}, 
broken time-reversal symmetry in the superconducting state has not yet been detected from muon-spin-relaxation measurements~\cite{Sundar2019PRB}.
Meanwhile, a recent surface impedance measurement suggests the presence of point nodes somewhere within the $ab$ plane \cite{Bae2020}.
The exact orientation of the gap nodes, however, has not yet been identified.
These issues need to be clarified from further careful experiments.

In this study, we have performed a field-angle-resolved measurement of the specific heat $C(T,H,\Omega)$ for UTe$_2$,
which is a powerful tool to identify the nodal structure~\cite{Sakakibara2016RPP,Sakakibara2007JPSJ,Kittaka2016JPSJ,Shimizu2016PRL,Shimizu2017PRB}. 
Low-energy quasiparticle excitations detected by $C(T,H,\Omega)$ support that 
the superconducting gap possesses point nodes in the $a$ direction alone.
Furthermore, unexpected features, reminiscent of the Pauli-paramagnetic effect, were observed in $\Hc2(T)$ and $C(H)$ under $H$ along the easy-magnetization $a$ axis,
although the Pauli-paramagnetic effect cannot destroy spin-triplet pairing with $\Vec{d} \perp \Vec{a}$ when $H \parallel a$. 
To solve this puzzle, we propose a vector order parameter $\Vec{d}(\Vec{k})=(\Vec{b} + i\Vec{c})(k_b + ik_c)$, 
whose $\Tc$ is tuned by easy-axis magnetization.

Single crystals of UTe$_2$ were grown by the chemical vapor transport method \cite{Ran2019Science}.
A single crystal with its mass of 5.9~mg weight was used in this study.
The directions of the orthorhombic axes of the sample were confirmed by single-crystal x-ray Laue photographs.
The specific heat was measured using the quasi-adiabatic heat-pulse method in a dilution refrigerator.
The addenda contribution was subtracted from the data shown below.
The magnetic field was generated by using a vector magnet, up to 7~T (3~T) along the horizontal $x$ (vertical $z$) direction.
By rotating the refrigerator around the $z$ axis using a stepper motor, the magnetic field direction was controlled three-dimensionally.


\begin{figure}[t]
\includegraphics[width=3.2in]{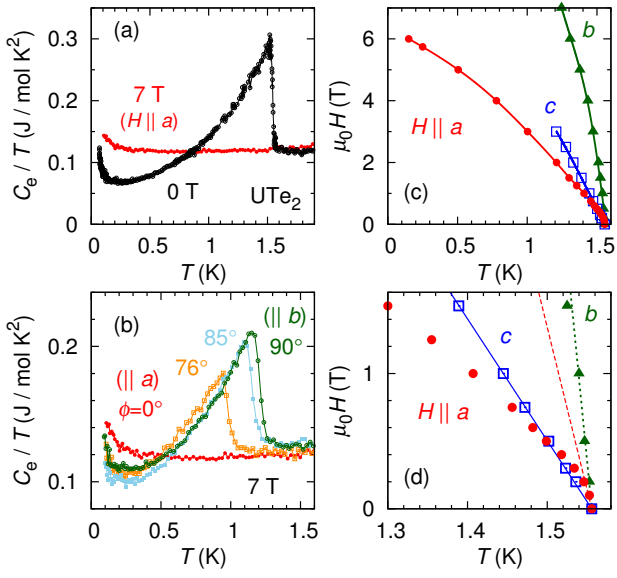} 
\caption{
(a) Temperature dependence of $C_{\rm e}/T$ at 0 and 7~T for $H \parallel a$.
(b) Temperature dependence of $C_{\rm e}/T$ at 7~T in various field orientations within the $ab$ plane.
(c) Field--temperature phase diagram for $H \parallel a$, $b$, and $c$ axes, 
and (d) its enlarged view near $\Tc$. 
Dashed, dotted, and solid lines in (d) represent initial slopes of $\Hc2(T)$ parallel to the $a$, $b$, and $c$ axes, respectively.
}
\label{CT}
\end{figure}

Figure \ref{CT}(a) plots $C_{\rm e}/T$ in zero field and in the normal state (at 7~T for $H \parallel a$) as a function of temperature.
Here, the phonon and nuclear contributions ($C_{\rm ph}$ and $C_{\rm N}$, respectively) are subtracted, i.e., $C_{\rm e}=C-C_{\rm ph}-C_{\rm N}$;
the Debye temperature is set to 125~K and $C_{\rm N}=0.135H^2/T^2$ $\muup$J/(mol K)
is obtained by using a nuclear spin Hamiltonian for $^{123}$Te and $^{125}$Te nuclei ($I=1/2$) with the natural abundances of 0.9\% and 7\%, respectively.
In zero field, a superconducting transition is observed at $\Tc=1.56$~K (onset).
The jump size is as large as the previous results~\cite{Aoki2019JPSJ,Metz2019PRB}, ensuring the high quality of the present sample.
Although a two-step specific-heat jump was recently reported~\cite{Hayes2020}, 
we checked eight crystals with varying quality using heat capacity, 
none of which showed a signature of such multiple transitions.

At low temperatures below 0.2~K, $C_{\rm e}/T$ shows a rapid upturn on cooling, as already reported~\cite{Metz2019PRB}. 
To satisfy the entropy-balance law, it is expected that the normal-state $C_{\rm e}/T$ is enhanced with decreasing temperature,
as proposed in Ref.~\onlinecite{Metz2019PRB}.
However, in the normal state at 7~T for $H \parallel a$, $C_{\rm e}/T$ does not show a substantial upturn at low temperatures.
This result suggests that the normal-state $C_{\rm e}/T$ varies with increasing $H$, as reported in high-field measurements~\cite{Imajo2019JPSJ}.

Figure~\ref{CT}(b) compares $C_{\rm e}(T)/T$ at 7~T in several field orientations within the $ab$ plane.
Here, the field angle $\phi$ denotes an azimuthal angle measured from the $a$ axis.
Even with the same magnetic field strength, 
the normal-state $C_{\rm e}/T$ at 1.5~K becomes larger with tilting $H$ away from the $a$ axis.
Furthermore, the entropy-balance law is not satisfied between the data at $\phi=0^\circ$ and $\phi \ne 0^\circ$.
These facts suggest that the normal-state $C_{\rm e}/T$ of UTe$_2$ depends not only on the field strength but also on its orientation. 

\begin{figure}[t]
\includegraphics[width=3.2in]{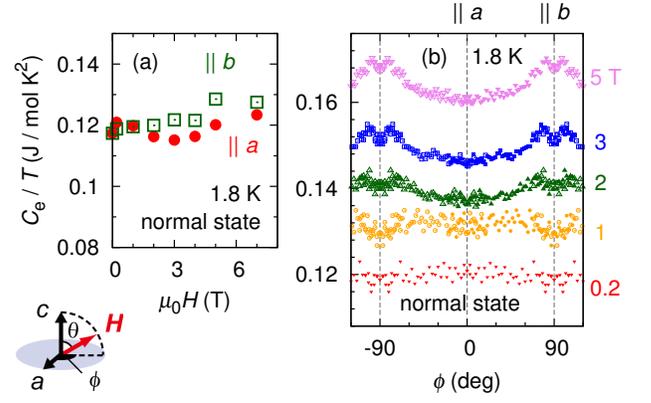} 
\caption{
(a) Field dependence of the normal-state $C_{\rm e}/T$ at 1.8 K for $H \parallel a$ and $b$.
(b) $C_{\rm e}/T$ at 1.8 K as a function of the azimuthal field angle $\phi$, taken under a rotating $H$ within the $ab$ plane, 
where the mirrored data with respect to symmetric axes are also plotted (open symbols). 
Each data set in (b) is vertically shifted by 0.01 J/mol K$^2$.
}
\label{normal}
\end{figure}

To examine the above possibility,
the effects of $H$ and its orientation on the normal-state $C_{\rm e}/T$ have been investigated as shown in Fig.~\ref{normal}. 
Indeed, the normal-state $C_{\rm e}/T$ changes with $H$ at 1.8~K ($>\Tc$) and shows a characteristic field-angle $\phi$ dependence under a rotating $H$ within the $ab$ plane.
An anomalous peak-dip-peak feature in $C_{\rm e}(\phi)$ becomes evident around $H \parallel b$ in the high-field region.
This feature may be related to longitudinal spin fluctuations along the $a$ axis 
because $C_{\rm e}(\phi)$ can be scaled approximately by the field component along the $a$ axis, $H_{\parallel a}=H\sin\theta\cos\phi$~\cite{KittakaSM}.
As depicted in Fig.~\ref{normal},
$\theta$ denotes a polar field angle measured from the $c$ axis.
While this abnormal normal-state behavior is in itself an intriguing and important issue, 
we leave it to future study and do not go into the detail here.
It is noted that this anomalous normal-state contribution is less important when $|\mu_0H_{\parallel a}|> 0.5$~T, 
where the longitudinal magnetic fluctuations are suppressed by a magnetic field, as discussed later.

The $H$--$T$ phase diagram of the present sample is shown in Fig.~\ref{CT}(c), 
which summarizes the onset temperature and onset field of superconductivity determined from $C_{\rm e}(T)$ and $C_{\rm e}(H)$ measurements.
The overall $\Hc2(T)$ behavior is consistent with the previous report from resistivity measurements~\cite{Aoki2019JPSJ}.
In this study, $\Hc2(T)$ near $\Tc$ is precisely determined from the thermodynamic measurements [Fig.~\ref{CT}(d)].
In sharp contrast to $\Hc2(T)$ for $H \parallel b$ and $c$,
$\Hc2(T)$ for $H \parallel a$ is clearly suppressed compared with the initial slope near $H \sim 0$~\cite{KittakaSM}.
A similar tendency of $\Hc2(T)$ can be found in magnetization measurements~\cite{Paulsen2020}.
The previous resistivity measurements~\cite{Aoki2019JPSJ} also support that 
$\Hc2(T)$ at low temperatures is more suppressed in $H \parallel a$ than in $H \parallel c$.
A possible origin of these unusual phenomena will be discussed later.

\begin{figure}[t]
\includegraphics[width=3.2in]{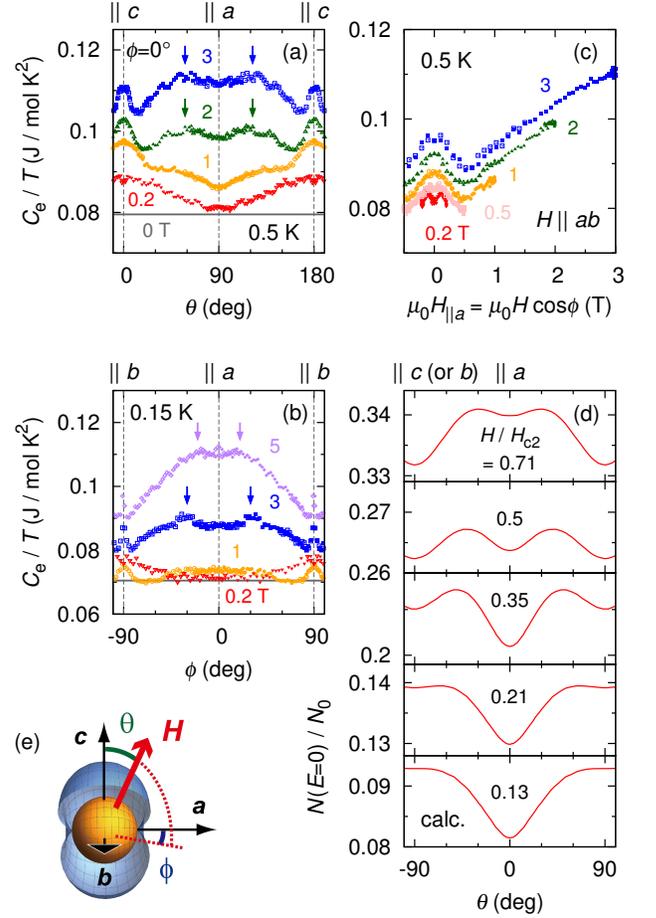} 
\caption{
Field-angle dependences of $C_{\rm e}/T$ under several magnetic fields rotated within (a) the $ac$ plane at 0.5 K and
(b) the $ab$ plane at 0.15~K.
Shoulder anomalies are indicated by arrows.
(c) The field-angle dependent $C_{\rm e}(\phi)/T$ under $H$ rotated within the $ab$ plane at 0.5~K 
plotted as a function of the $a$-axis component of $H$.
In these figures, the mirrored data with respect to symmetric axes are also plotted (open symbols). 
Numbers labeling the curves represent the magnetic field $\mu_0H$ in tesla.
(d) Calculated results of $N(E=0)$ normalized by $N_0$ for an axial state with two point nodes as a function of field angle, 
where $\theta=0^\circ$ is the direction of point nodes (taken from Ref.~\onlinecite{Tsutsumi2016PRB}).
(e) The gap structure possessing point nodes along the $a$ direction.
}
\label{Cphi}
\end{figure}

To investigate the gap anisotropy of UTe$_2$, 
the field-angle dependence of $C_{\rm e}/T$ has been measured in a rotating $H$ within the $ab$, $ac$, and $bc$ planes;
the results are presented in Figs.~\ref{Cphi}(a), \ref{Cphi}(b), and the Supplemental Material~\cite{KittakaSM}.
It is noted that anomalous peaks are observed 
at $\theta=0^\circ$ and $180^\circ$ ($\phi=\pm 90^\circ$) in Fig.~\ref{Cphi}(a) [Fig.~\ref{Cphi}(b)], 
whose widths become narrower with increasing $H$.
By contrast, such a sharp peak does not appear around $H \parallel a$.
These dip-peak-dip features around $H \perp a$ are qualitatively similar to the peak-dip-peak feature observed at 1.8~K shown in Fig. \ref{normal}(b),
though the sign of the oscillation is reversed.
Plausibly, these anomalies are also related to Ising-type spin fluctuations that are easily suppressed by $H_{\parallel a}$.

Figure~\ref{Cphi}(c) plots $C_{\rm e}(\phi)/T$ at 0.5~K measured under a rotating field within the $ab$ plane as a function of $\mu_0H_{\parallel a}$.
The dip-peak-dip structure can be scaled clearly by $H_{\parallel a}$, and the dips are located at $|\mu_0H_{\parallel a}|\sim0.5$~T.
This dip-peak-dip structure appearing for $|\mu_0H_{\parallel a}|<0.5$~T disturbs the detection of low-energy quasiparticle excitations reflecting gap anisotropy.
In other words, when $|\mu_0H_{\parallel a}|>0.5$~T, the data are expected to be dominated by quasiparticle excitations.
Accordingly, to discuss the gap nodal structures, 
we restrict attention to the field orientation range well away from the plane perpendicular to the $a$ axis
and in the magnetic field range for $\mu_0H > 1$~T.

In Fig.~\ref{Cphi}(a), at 0.5 K and a low field of 0.2~T, 
a local minimum exists in $H \parallel a$ ($\phi=0^\circ$, $\theta=90^\circ$),
which is probably dominated by the dip-peak-dip structure.
However, with increasing $\mu_0H$ above 1~T, 
in which the data are less affected by the dip-peak-dip anomaly for $|\theta-90^\circ|\lesssim60^\circ$, 
$C_{\rm e}(\phi,\theta=90^\circ)$ remains to have a local minimum around $H \parallel a$ 
and exhibits a shoulder structure 
at intermediate-field angles slightly away from the $a$ axis, as indicated by arrows.
These features would mainly reflect the anisotropy of low-energy quasiparticle excitations.
A qualitatively similar feature can be observed in $C_{\rm e}(\phi=0^\circ,\theta)$ at 0.15~K as well [see Fig.~\ref{Cphi}(b)].

\begin{figure}[t]
\includegraphics[width=3.2in]{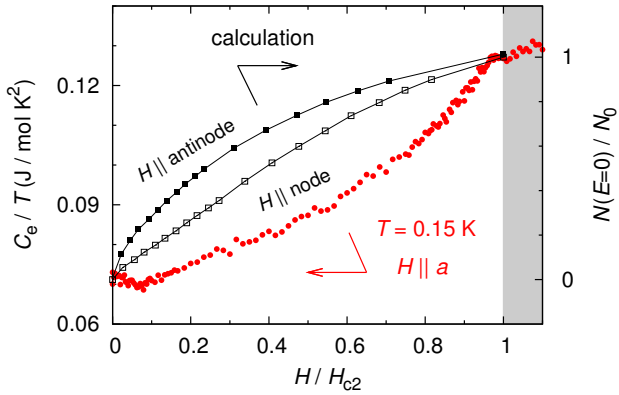} 
\caption{
The specific-heat data $C_{\rm e}(H)/T$ at 0.15~K as a function of $H/\Hc2$ for $H \parallel a$ (circles), where $\mu_0\Hc2$ is 6~T. 
Open (solid) squares are $N(E=0)/N_0$ calculated for an axial state with two point nodes 
under $H$ parallel (perpendicular) to the nodal direction (taken from Ref.~\onlinecite{Miranovic2003PRB}).
}
\label{CH}
\end{figure}

Figure \ref{CH} represents $C_{\rm e}(H)/T$ at 0.15~K as a function of $H/\Hc2$ along the $a$ direction. 
Whereas there exist various experimental results supporting the presence of nodes somewhere in the gap, 
$C_{\rm e}(H)$ for $H \parallel a$ does not show a rapid increase in the low-field region at any temperature~\cite{KittakaSM}.
This result may suggest the absence of nodal quasiparticle excitations in this field orientation.
However, there also remains a possibility that low-energy quasiparticle excitations are masked by 
a significant contribution from the dip-peak-dip anomaly for $|\mu_0H_{\parallel a}|<0.5$~T.

In the high-field superconducting region for $H \parallel a$, 
where the contribution from the dip-peak-dip anomaly becomes small,
$C_{\rm e}(H)$ shows a convex downward curvature with increasing $H$ at 0.15~K. 
This feature is \textit{apparently} similar to the Pauli-paramagnetic effect 
which breaks Cooper pairs to make spins polarized along the field orientation~\cite{Ichioka2007PRB,Machida2008PRB}.
However, the Pauli-paramagnetic effect would not be allowed for UTe$_2$ in $H \parallel a$ 
because the $a$ direction corresponds to the easy-magnetization axis and spins of triplet Cooper pairs ($\Vec{d} \perp \Vec{a}$) can be polarized in this direction.
Therefore, an unusual mechanism of spin-triplet superconductivity is required for UTe$_2$. 

Let us discuss the gap symmetry of UTe$_2$.
On theoretical grounds, the low-temperature specific heat is proportional to the zero-energy quasiparticle density of states $N(E=0)$.
The field and field-angle dependences of $N(E=0)$ calculated for a point-nodal superconductor were already reported~\cite{Miranovic2003PRB,Tsutsumi2016PRB}. 
The present observations in $C(T,H,\Omega)$, except for anomalous peaks in its angular dependence around $H \perp a$, 
are in good agreement with the calculated results based on a microscopic theory 
assuming the presence of linear point nodes in the gap along the $a$ direction \cite{Tsutsumi2016PRB},
as presented in Fig.~\ref{Cphi}(d).
In particular, the appearance of a shoulderlike anomaly 
with a local minimum along the nodal direction 
in intermediate fields [Figs.~\ref{Cphi}(a) and \ref{Cphi}(b)] is also consistent with the theoretical calculations [Fig.~\ref{Cphi}(d)]. 
Note that the agreement between the model calculations and the experiment is qualitative, because the model calculations ignore the detailed band structures as well as  the $\Hc2$ anisotropy.  
As represented by squares in Fig.~\ref{CH},
theoretical calculations for the axial state with two point nodes along the $a$ axis predict the gradual $H$-linear increase of $N(E=0)$ in the point-node direction,
in sharp contrast to the steep increase in antinodal directions~\cite{Miranovic2003PRB}.
Favorably, this prediction is also compatible with the present experimental observations.
Because the presence of linear point nodes has been indicated in previous reports~\cite{Ran2019Science,Nakamine2019JPSJ,Metz2019PRB},
the present results, supporting their orientation along the $a$ direction [Fig.~\ref{Cphi}(e)],
lead to an indication that the orbital part of the order parameter for UTe$_2$ is a chiral state $k_b + ik_c$ or a helical state $k_b\Vec{c}+k_c\Vec{b}$ 
belonging to the $B_{3u}$ representation classified for strong spin-orbit coupling \cite{Machida2001PRL,Annett1990AP,Ishizuka2019PRL,Nevidomskyy2020}.
The former is consistent with a chiral $p+ip$-type pairing concluded from STM experiments~\cite{Jiao2019}.

Regarding a possible mechanism of anomalous behaviors in $\Hc2(T)$ and $C_{\rm e}(H)$ for $H \parallel a$,
we here consider a phenomenological model based on the Ginzburg-Landau framework in which the degeneracy of nonunitary order parameters 
$\Vec{d} \propto (\Vec{b} \pm i\Vec{c})$ with equal spin pairing (i.e., $\Delta_{\uparrow\uparrow}$ and $\Delta_{\downarrow\downarrow}$)
is lifted by the easy-axis magnetization $M_a$;
one of the order parameters ($\Delta_{\uparrow\uparrow}$) arises at $\Tc$ and
the other ($\Delta_{\downarrow\downarrow}$) appears at a lower temperature.
In this model, $\Tc$ of $\Delta_{\uparrow\uparrow}$ is written as $\Tc(M)=\T0 + \eta M_a$~\cite{Machida2001PRL,Machida2020JPSJ,Machida2020JPSJ2}.
Here, $\eta$ is a positive constant coefficient.
In general, $M_a$ has a nonlinear component of $H$. 
Therefore, we can reasonably assume $M_a(H,T) \sim M_0(T) + \alpha(T) H-\beta(T) H^2$ at low fields by using positive coefficients $\alpha$ and $\beta$.
The spontaneous magnetization or the root-mean-square average of longitudinal magnetization fluctuations $M_0$ breaks the degeneracy of $\Delta_{\uparrow\uparrow}$ and $\Delta_{\downarrow\downarrow}$ in zero field.
Then, we obtain $\Tc(M_a)=\T0^\ast+\eta [\alpha(T) H-\beta(T) H^2]$, where $\T0^\ast=\T0+\eta M_0$ ($\sim1.6$~K).
From this equation, it is suggested that the slope of $\Hc2(T)$ is enhanced when $H$ is sufficiently low~\cite{KittakaSM},
because of $\Hc2(T) \approx \zeta[\Tc(M_a)-T]$ ($\zeta$ is a positive coefficient),
but the slope is suppressed at higher fields due to the nonlinear term in $M_a(H)$.
Indeed, the slope of $\Hc2(T)$ in $H \parallel a$ for UTe$_2$ becomes small at low temperatures (in high fields) [see Fig.~\ref{CT}(d)].
A similar behavior in $\Hc2(T)$ was also reported for a reentrant superconductor URhGe along the magnetic-easy-axis direction 
in the lower-field superconducting phase~\cite{Hardy2005PRL}. 
Furthermore, the convex downward curvature in the low-temperature $C_{\rm e}(H)$ for $H \parallel a$ (Fig.~\ref{CH}) can also be explained qualitatively by this model;
if we assume $C_{\rm e}(H)/C_{\rm e}(H=0) \sim H/\Hc2(H)$ for the field direction parallel to the point nodes,
$C_{\rm e}(H)/C_{\rm e}(H=0) \sim H/\zeta[\Tc^\ast+\eta(\alpha H -\beta H^2)]$.
Under $H$ along hard-magnetization axes, these unusual phenomena are not expected
because $M_a$ does not change significantly with $H$.
Thus, the present study may capture the universal nature of nonunitary equal-spin-triplet superconductivity. 

On the basis of these results, the order parameter $(\Vec{b} + i\Vec{c})(k_b + ik_c)$ is a leading candidate for the superconductivity in UTe$_2$.
In this case, a secondary superconducting transition is expected below $\Tc$,
which was recently suggested by a sudden drop of $1/T_1$ around $T \sim 0.15$~K~\cite{Nakamine2019JPSJ}.
Unfortunately, the specific heat at an ambient pressure shows an unusual enhancement at low temperatures,
hindering a possible weak anomaly associated with this second transition.
However, a two-step specific-heat anomaly was found under hydrostatic pressure~\cite{Daniel2019ComPhys,Aoki2020JPSJ},
suggesting an occurrence of multiple superconducting phases. 
In order to lift the degeneracy of the multiple order parameters in zero field, a spontaneous magnetization or very slow longitudinal spin fluctuations are needed. 
This requirement suggests a possibility that a short-range magnetic order 
of $M_a$ developing in UTe$_2$ above $\Tc$~\cite{Ran2019Science,Miyake2019JPSJ,Tokunaga2019JPSJ,Sundar2019PRB} breaks the degenerate order parameters; 
this is similar to the case of UPt$_3$ \cite{Machida1989JPSJ,Machida1991PRL} which shows a double superconducting transition coupled with a short-range antiferromagnetic order \cite{Hayden1992PRB,Trappmann1991PRB}.
One may suspect that the small $\gamma_0$, less than half of the normal-state value~\cite{Aoki2020JPSCP}, is inconsistent with the nonunitary pairing scenario.
However, this may happen if the majority-spin bands have a larger effective mass than minority-spin bands;
$\gamma_0/\gamma_{\rm n}=50\%$ is not the universal number for the nonunitary superconductors.
Although the proposed gap symmetry is not classified by group theory in $D_{2h}$,
a chiral vector $\Vec{l}$ pointing to the magnetic-easy axis may stabilize the proposed pairing via the energy of $\Vec{l} \cdot \Vec{M}$.
In the weak spin-orbit coupling case, SO(3) symmetry allows the $\Vec{b} + i\Vec{c}$ state~\cite{Machida2001PRL,Annett1990AP}.

In summary, we have performed field-angle-resolved measurements of the specific heat on UTe$_2$.
Our results, in particular, the characteristic field evolution in $C_{\rm e}(\Omega)$, 
suggest that linear point nodes are located along the $a$ direction in the superconducting gap.
Based on this gap structure,
the orbital part of the order parameter can be characterized by a chiral $p$-wave form $k_b + ik_c$ or a helical state $k_b\Vec{c}+k_c\Vec{b}$.
Furthermore, unusual $\Hc2(T)$ and $C_{\rm e}(H)$ behaviors have been found under $H$ along the easy-magnetization $a$ axis,
which can be explained by a phenomenological model for a nonunitary equal-spin-triplet pairing tuned by the easy-axis magnetization.
On the basis of these findings, together with recent STM results~\cite{Jiao2019}, the vector order parameter $\Vec{d}(\Vec{k})=(\Vec{b} + i\Vec{c})(k_b + ik_c)$ is a leading candidate for UTe$_2$.
Further experimental investigations are needed to validate these nonunitary superconducting order parameters.

\begin{acknowledgments}
We thank J. Flouquet, J. P. Brison, G. Knebel, and K. Ishida for fruitful discussion.
We are thankful for all the support from Institute for Materials Research, Tohoku University in growing single-crystalline uranium-based samples using the joint research facility at Oarai.
This work was supported by a Grant-in-Aid for Scientific Research on Innovative Areas ``J-Physics'' (JP15H05883, JP18H04306, JP15H05882, JP15H05884, JP15K21732) from MEXT,
and KAKENHI (JP18H01161, JP17K05553, JP19H00646, JP15H0574, JP17K14328) from JSPS.
\end{acknowledgments}

\bibliography{C:/usr/local/share/texmf/bibref/ref_UTe2.bib}

\clearpage
\onecolumngrid
\appendix
\begin{center}
{\large Supplemental Material for \\
\textbf{Orientation of point nodes and nonunitary triplet pairing tuned by the easy-axis magnetization\\ in UTe$_2$}}\\
\vspace{0.1in}
Shunichiro Kittaka,$^{1,2}$ Yusei Shimizu,$^{3}$ Toshiro Sakakibara,$^{1}$ Ai Nakamura,$^{3}$ Dexin Li,$^{3}$ Yoshiya Homma,$^{3}$ Fuminori Honda,$^{3}$ Dai Aoki,$^{3}$ and Kazushige Machida,$^{4}$\\
{\small 
\textit{$^1$Institute for Solid State Physics, University of Tokyo, Kashiwa, Chiba 277-8581, Japan}\\
\textit{$^2$Department of Physics, Chuo University, Kasuga, Bunkyo-ku, Tokyo 112-8551, Japan}\\
\textit{$^3$IMR, Tohoku University, Oarai, Ibaraki 311-1313, Japan}\\
\textit{$^4$Department of Physics, Ritsumeikan University, Kusatsu, Shiga 525-8577, Japan}\\
}
(Dated: \today)
\end{center}

\section{I. Anomalous behaviors in specific heat}
Figures \ref{CH_SM}(a)-\ref{CH_SM}(c) show the field dependences of the specific-heat data $C_{\rm e}/T$ of UTe$_2$ in three field orientations parallel to the $a$, $b$, and $c$ axes at several temperatures.
An anomaly at $\Hc2$ can be detected only for $H \parallel a$ because of the limit of our measurement system.
With decreasing temperature as low as 0.15 K, a prominent peak develops around 0.2 T for $H \parallel b$ and $H \parallel c$. 
This anomaly may be related to the abnormal behavior in $C_{\rm e}/T$ at low temperatures, 
which breaks the entropy-balance law. 
Unfortunately, this anomaly disturbs investigation of low-energy quasiparticle excitations at low temperatures.

Figures \ref{Cphi_SM}(a) and \ref{Cphi_SM}(b) show the field-angle $\phi$ dependence of $C_{\rm e}/T$ at 0.15 and 0.5 K, respectively, in a rotating magnetic field within the $ab$ plane.
At high fields above 5~T, a specific-heat jump at $\Hc2$ is clearly detected in $C_{\rm e}(\phi)$.
Furthermore, anomalous dip-peak-dip structure is observed around $H \parallel b$ and $H \parallel c$; 
the peak develops and becomes sharper with increasing $H$.
At 0.15~K, although the $\Hc2$ anomaly becomes small due to the suppression of the specific-heat jump at $\Tc(H)$,
the dip-peak-dip anomaly remains observed clearly.
Such a peak anomaly does not arise around $H \parallel a$.
Figure \ref{bc_SM} shows the field angle $\theta$ dependence of $C_{\rm e}/T$ at 0.5 K in a rotating magnetic field within the $bc$ plane.
With increasing magnetic field, its angle dependence is unchanged qualitatively.

Figures~\ref{scale_SM}(a)-\ref{scale_SM}(c) plot $C_{\rm e}(\phi)/T$ at 0.15, 0.5, and 1.8 K, respectively, 
as a function of the $a$-axis component of the magnetic field, i.e., $H_{\parallel a}=H\sin\theta\cos\phi$, at $\theta=90^\circ$.
A sharp peak or dip exists in $C_{\rm e}(H_{\parallel a})$ centered at $H_{\parallel a}=0$ whose width is robust against the magnetic-field strength at any temperature.
These results suggest that the specific heat in both normal and superconducting states is affected by longitudinal spin fluctuations along the $a$ axis,
which can be easily suppressed by $H_{\parallel a}$.

\section{II. Field orientation effect on $\Tc$ at low fields}
Figure \ref{CT_SM} compares the temperature dependences of $C_{\rm e}/T$ at 0.2~T for $H \parallel a$ and $H \parallel c$.
Although the low-temperature $\Hc2$ is smaller in $H \parallel a$ than in $H \parallel c$,
$\Tc(H)$ at 0.2~T is higher in $H \parallel a$.
This result demonstrates that the initial slope of $\Hc2(T)$ near $\Tc$ is larger in $H \parallel a$.
By contrast, at a slightly higher magnetic field of 0.75~T, $\Tc(H)$ becomes lower in $H \parallel a$.
These features clearly evidence that $\Hc2(T)$ is suppressed on cooling compared with its initial slope at $H \sim 0$ for $H \parallel a$.

\section{III. Phenomenological model for nonunitary spin-triplet pairing}

In general, spin-triplet pairing ($S=1$) has spin degrees of freedom. 
Therefore, its multiple order parameters can be coupled with a magnetization.
Here, we consider the case of nonunitary order parameters $\Delta_{\uparrow\uparrow}$ and $\Delta_{\downarrow\downarrow}$,
whose spins can be polarized along the easy-magnetization axis.
If there exists a spontaneous magnetization or the root-mean-square average of longitudinal spin fluctuations along the easy-magnetization axis ($M_a$), 
the degeneracy of $\Delta_{\uparrow\uparrow}$ and $\Delta_{\downarrow\downarrow}$ is lifted and their transition temperatures are split.
In this case, an onset $\Tc$ can be described as
\begin{align}
\Tc(M_a)=T_{\rm c0} + \eta M_a
\end{align}
with a positive constant coefficient $\eta$.
In general, the low-temperature $M_a$ roughly behaves as
\begin{align}
M_a(H,T) \sim M_0^\ast(H)-b(H)T^2. 
\end{align}
Here, $M_0^\ast$ is the temperature-independent part of $M_a$.
Because the temperature dependence of the upper critical field can be simply described as 
\begin{align}
\Hc2(T) \sim \zeta[\Tc(M_a)-T], 
\end{align}
the slope of $\Hc2(T)$ near $H \sim 0$ can be expressed as
\begin{align}
\frac{d\Hc2}{dT}&=\zeta\frac{d \Tc}{d M_a}\frac{d M_a}{d T}-\zeta\\
&=-2\eta\zeta b(H) T-\zeta. \label{Hc2slope}
\end{align}
The first term in the right-hand side of eq.~\eqref{Hc2slope} reinforces the initial slope of $\Hc2(T)$.

When we apply the magnetic field along the easy-magnetization axis, $M_a$ usually changes significantly with a nonlinear term at low fields; e.g., 
\begin{align}
M_a(H) \sim M_0(T)+\alpha(T)H-\beta(T)H^2.
\end{align}
Then, $\Tc(M_a)$ changes as
\begin{align}
\Tc(M_a)=T_{\rm c0}^\ast + \eta \alpha H -\eta \beta H^2, \label{eqTc}
\end{align}
where $T_{\rm c0}^\ast=T_{\rm c0}+\eta M_0$.
The second term in the right-hand side of eq.~\eqref{eqTc} contributes to the increase of $\Tc$;
if it is dominantly large, the initial slope of $\Hc2(T)$ near $H \sim 0$ can even become positive.
When the third term develops in the high-field region, the enhancement of $\Tc$ is suppressed,
and a relatively small slope can occur in $\Hc2(T)$ at low temperatures in high fields.
These features qualitatively match the experimental observations for UTe$_2$.

\setcounter{figure}{0}
\renewcommand{\thefigure}{S\arabic{figure}}

\begin{figure}[h]
\includegraphics[width=5.in]{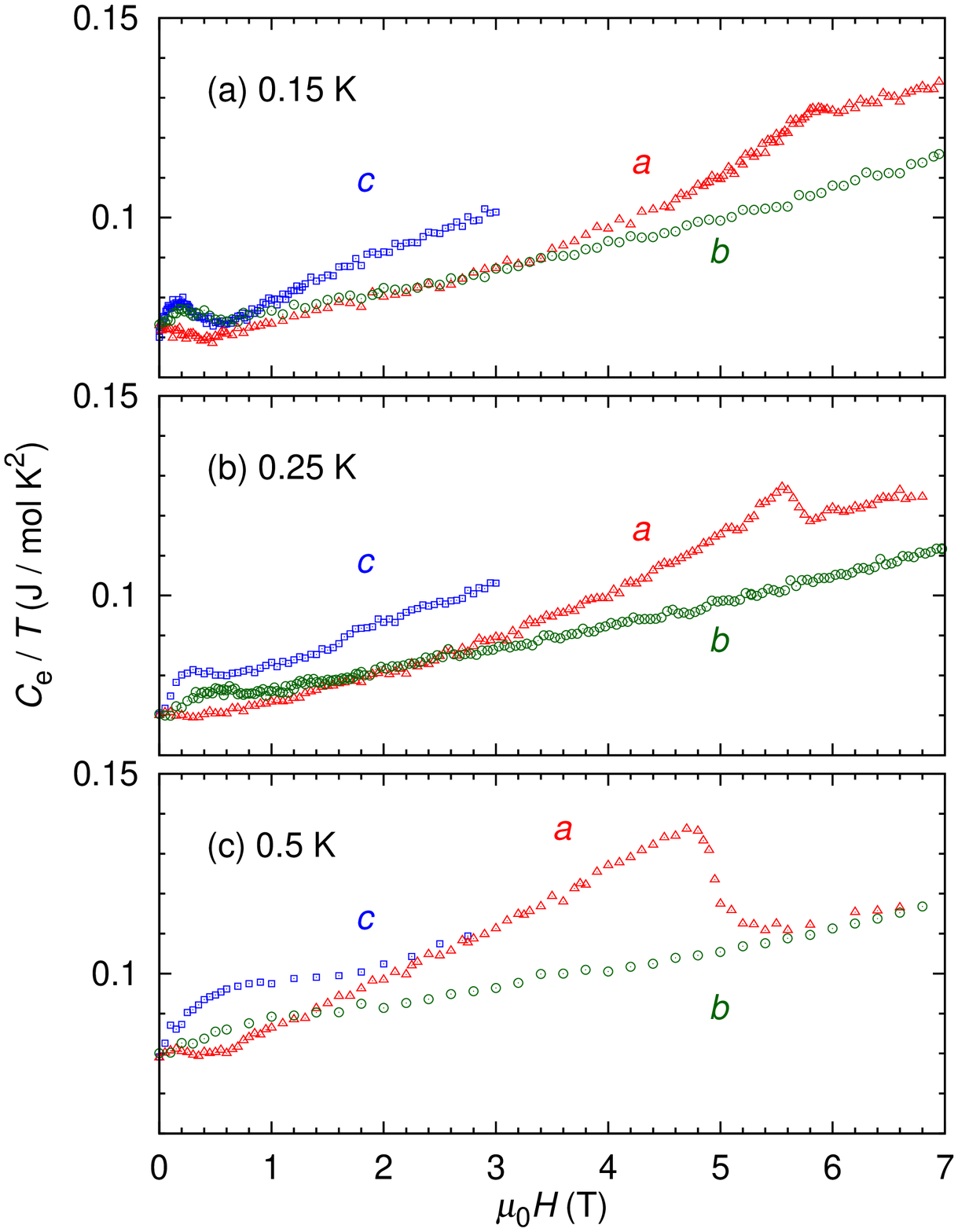} 
\caption{
Field dependence of $C_{\rm e}/T$ at (a) 0.15, (b) 0.25, and (c) 0.5~K in three orientations along the $a$, $b$, and $c$ axes.
}
\label{CH_SM}
\end{figure}

\begin{figure}[h]
\includegraphics[width=6.in]{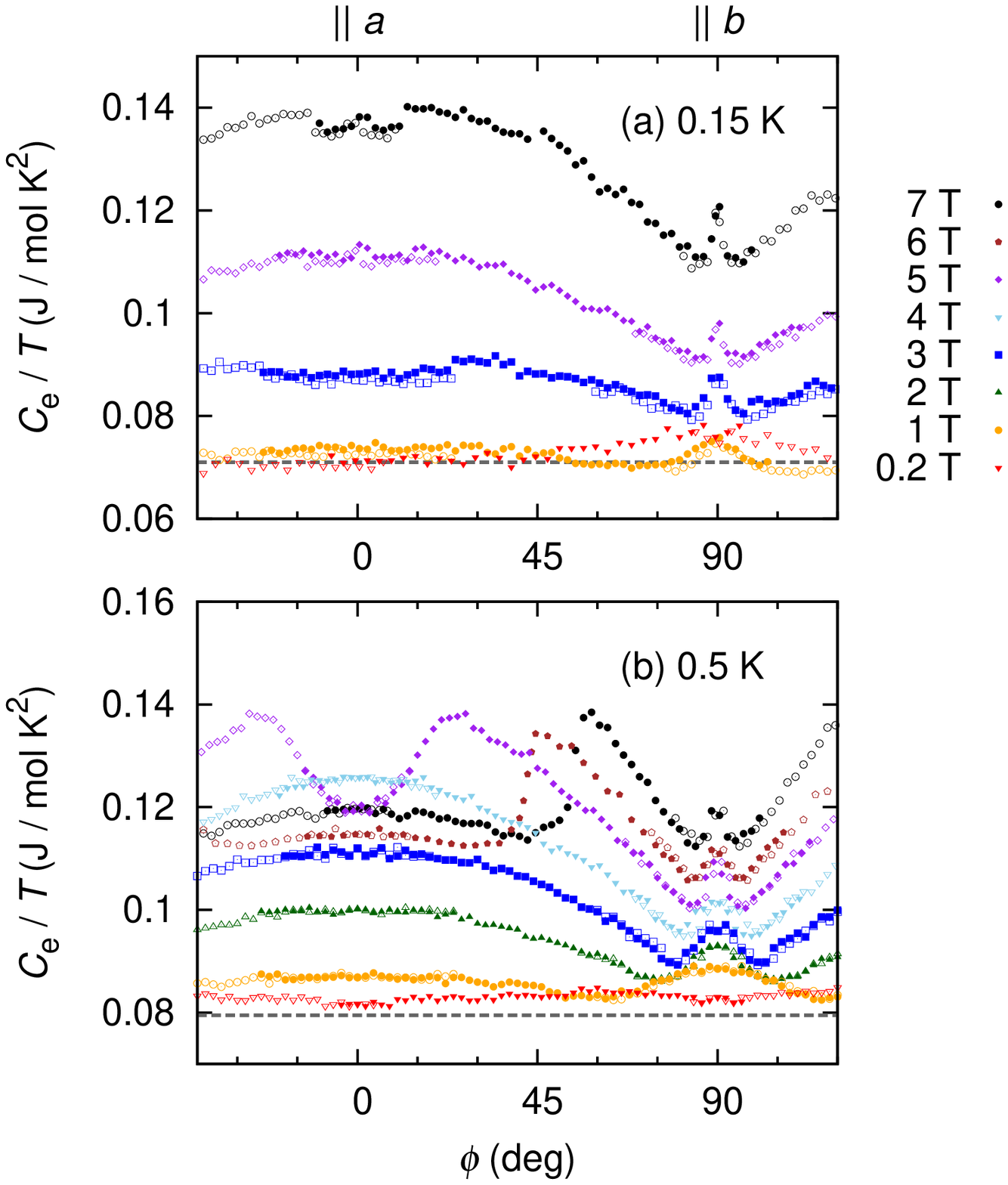} 
\caption{
Field-angle $\phi$ dependence of $C_{\rm e}/T$ at (a) 0.15 and (b) 0.5~K under several magnetic fields rotated within the $ab$ plane, 
where $\phi$ is the field angle measured from the $a$ axis.
In these figures, the mirrored data with respect to symmetric axes are also plotted (open symbols).
}
\label{Cphi_SM}
\end{figure}

\begin{figure}[h]
\includegraphics[width=6.in]{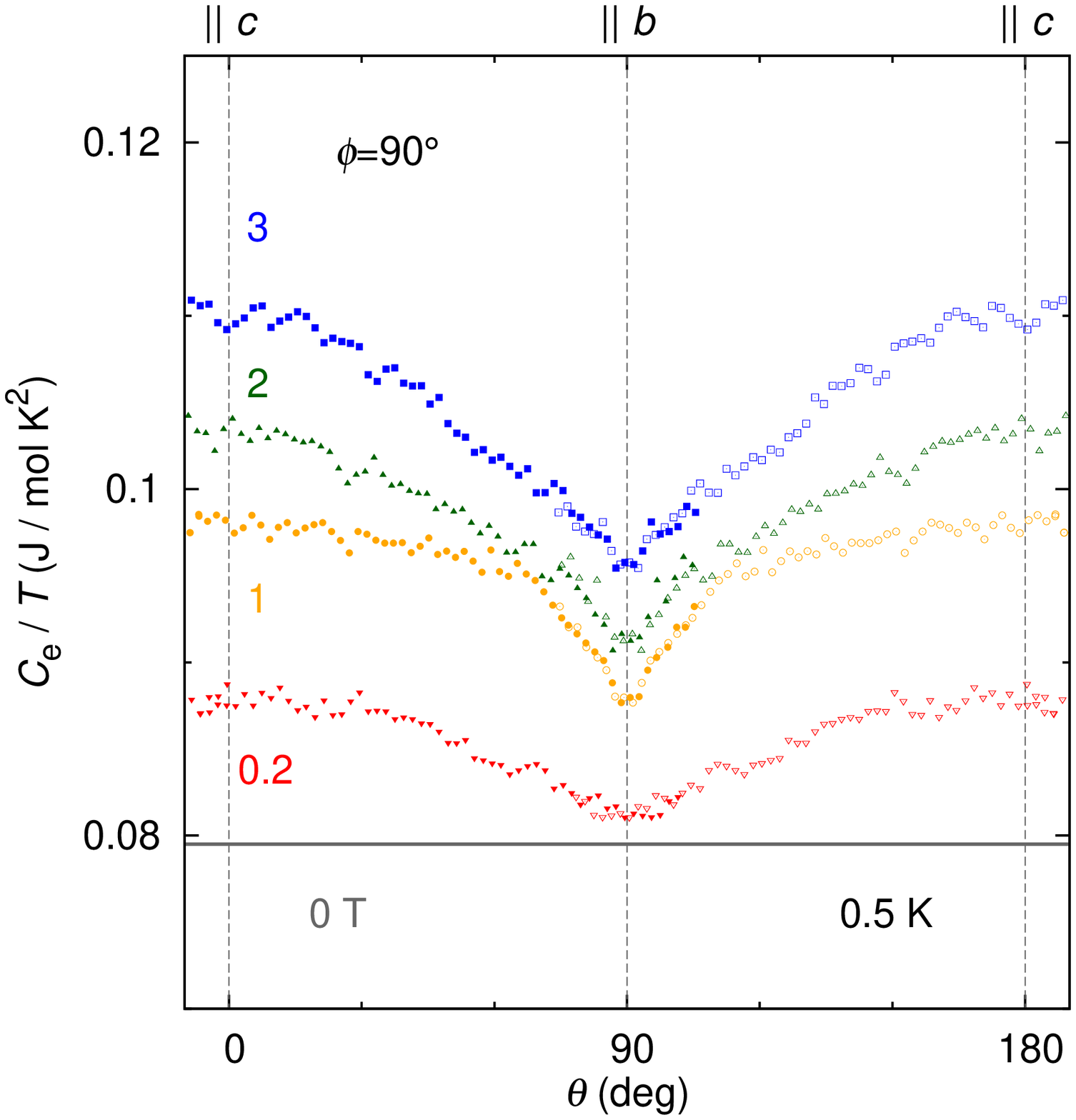} 
\caption{
Field-angle $\theta$ dependence of $C_{\rm e}/T$ at 0.5~K under several magnetic fields rotated within the $bc$ plane,
where the mirrored data with respect to $\theta=90^\circ$ are also plotted (open symbols). 
}
\label{bc_SM}
\end{figure}

\begin{figure}[h]
\includegraphics[width=6.in]{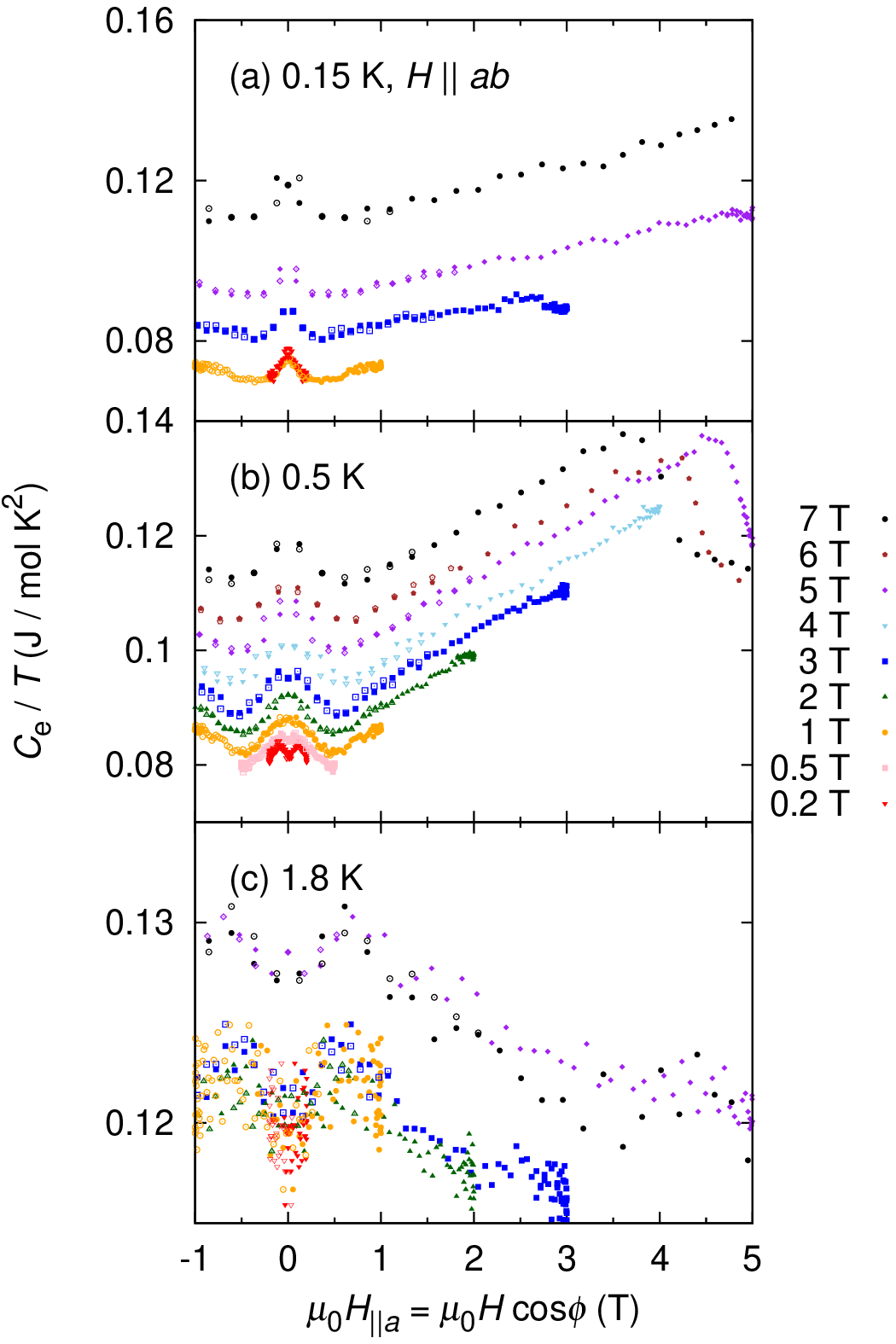} 
\caption{
$C_{\rm e}(\phi)/T$ at (a) 0.15, (b) 0.5, and (c) 1.8~K under several magnetic fields rotated within the $ab$ plane ($\theta=90^\circ$)
plotted as a function of the $a$-axis component of the magnetic field $\mu_0H_{\parallel a}=\mu_0H\cos\phi$. 
In these figures, the mirrored data with respect to $H_{\parallel a}=0$ are also plotted (open symbols).
}
\label{scale_SM}
\end{figure}

\begin{figure}[h]
\includegraphics[width=5.in]{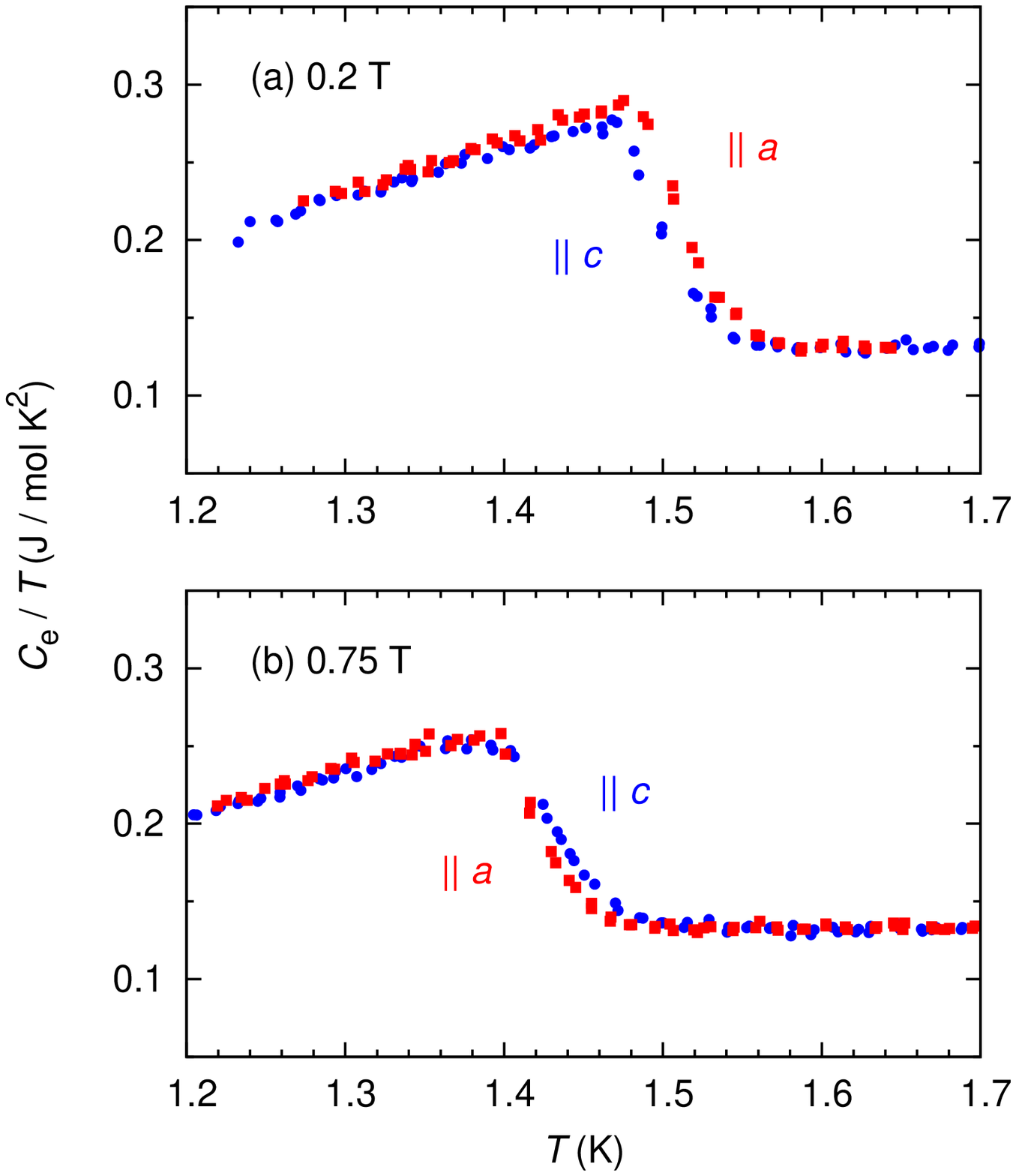} 
\caption{
Temperature dependence of $C_{\rm e}/T$ at (a) 0.2 and (b) 0.75~T for $H \parallel a$ and $H \parallel c$.
}
\label{CT_SM}
\end{figure}

\end{document}